\documentclass[fleqn,usenatbib]{mnras}

\usepackage{newtxtext,newtxmath}

\usepackage[T1]{fontenc}

\DeclareRobustCommand{\VAN}[3]{#2}
\let\VANthebibliography\thebibliography
\def\thebibliography{\DeclareRobustCommand{\VAN}[3]{##3}\VANthebibliography}

\usepackage{graphicx}	
\usepackage{amsmath}	
\usepackage[dvipsnames]{xcolor}

\newcommand{\hMpc}{h^{-1}\mathrm{Mpc}}
\newcommand{\hGpc}{h^{-1}\mathrm{Gpc}}

\title[Approximate lightcones in spherical shells]{Solving small-scale clustering problems in
approximate lightcone mocks}

\author[A. Smith et al.]{
Alex Smith$^{1,2,3}$\thanks{E-mail: alex.smith@ed.ac.uk},
Shaun Cole$^{1}$,
Cameron Grove$^{1}$,
Peder Norberg$^{1,4}$,
and Pauline Zarrouk$^{5}$
\vspace*{4pt} \\ 
\scriptsize $^{1}$ Institute for Computational Cosmology, Department of Physics, Durham University, South Road, Durham DH1 3LE, UK\vspace*{-2pt}\\
\scriptsize $^{2}$ IRFU, CEA, Universit\'e Paris-Saclay, F-91191 Gif-sur-Yvette, France\vspace*{-2pt} \\
\scriptsize $^{3}$ Institute for Astronomy, University of Edinburgh, Royal Observatory, Blackford Hill, Edinburgh EH9 3HJ, UK\vspace*{-2pt}\\
\scriptsize $^{4}$ Centre for Extragalactic Astronomy, Department of Physics, Durham University, South Road, Durham DH1 3LE, UK \vspace*{-2pt} \\
\scriptsize $^{5}$ Sorbonne Universit\'e, Universit\'e Paris Diderot, CNRS/IN2P3, Laboratoire de Physique Nucl\'eaire et de Hautes Energies, LPNHE, 4 place Jussieu, F-75252 Paris, France \vspace*{-2pt} \\
}

\date{Accepted XXX. Received YYY; in original form ZZZ}

\pubyear{2022}

\begin{document}
\label{firstpage}
\pagerange{\pageref{firstpage}--\pageref{lastpage}}
\maketitle

\begin{abstract}
Realistic lightcone mocks are important in the clustering analyses of large 
galaxy surveys. For simulations where only the snapshots are available, it is 
common to create approximate lightcones by joining together the snapshots in 
spherical shells. We assess the two-point clustering measurements of central galaxies in approximate
lightcones built from the Millennium-XXL simulation, which are constructed using different numbers of snapshots.
The monopole and quadrupole of the real-space correlation function is strongly boosted on small
scales below $1~\hMpc$, due to some galaxies being duplicated at the boundaries between snapshots in the
lightcone. When more snapshots are used, the total number of duplicated galaxies 
is approximately constant,
but they are pushed to smaller separations. The effect of this in redshift
space is small, as long as the snapshots are cut into shells in real space. Randomly
removing duplicated galaxies is able to reduce the excess clustering signal. Including
satellite galaxies will reduce the impact of the duplicates, since many small-scale pairs
come from satellites in the same halo. Galaxies that are missing from the lightcone at the boundaries 
can be added to the lightcone by having a small overlap between each shell. This effect will impact analyses that use very small-scale clustering measurements, and when using mocks to test the impact of fibre collisions.
\end{abstract}




\section{Introduction}

The use of realistic mock galaxy catalogues is an essential requirement
in the clustering analysis of galaxies in large galaxy surveys.
In order to make redshift space distortion \citep[RSD;][]{Kaiser1987} and 
Baryon Acoustic Oscillation \citep[BAO; e.g.][]{Cole2005,Eisenstein2005} 
measurements from the two-point clustering statistics, the theoretical models must be validated
on mock catalogues, where the underlying cosmology is known. Additionally,
the use of mocks allows the systematics affecting these measurements
to be quantified, and for the development of methods to mitigate them
\citep[e.g.][]{Smith2019,Sugiyama2020,DeRose2021}.
These allow us to place constraints on theories of dark energy, and
theories of modified gravity \citep[e.g.][]{Guzzo2008}.

Current and future galaxy surveys, such as the Dark Energy Spectroscopic 
Instrument \citep[DESI;][]{DESI2016science,DESI2016instrument, DESI2022}, 
Legacy Survey of Space and Time \citep[LSST;][]{Ivezic2019},
Roman Space Telescope \citep{Spergel2015},
and Euclid \citep{Laureijs2011}, will map many millions
of galaxies. In order to reach the required high precision BAO and RSD measurements, it is essential that the mock catalogues used are as accurate as possible.

Mock galaxy catalogues are constructed using the outputs of N-body simulations.
Due to the large volumes required, dark-matter-only simulations are typically used,
which are then populated with galaxies. There are several methods which are commonly
used to add galaxies to the dark matter haloes, such as the halo occupation distribution
\citep[HOD; e.g.][]{Smith2017,Alam2020,Smith2020,Alam2021,Rossi2021}, where the probability a halo contains central and satellite
galaxies depends on the halo mass. Subhalo abundance matching \citep[SHAM; e.g.][]{Rodriguez2016,Safonova2021}
ranks the subhaloes in the simulation based on a property (e.g. circular velocity),
placing the brightest galaxies in the most massive subhaloes (with scatter).
Semi-analytic models \citep[SAM; e.g.][]{Cole2000,Benson2010} describe the physics of galaxy formation and evolution,
using analytic techniques.

To emulate an observed dataset, it is necessary to create lightcone mocks, where the
galaxy properties evolve with the distance to the observer. Distant haloes or galaxies in the lightcone 
are output at early times in the simulation, and nearby galaxies at late times.
Ideally, a direct lightcone output of the simulation would be used, where particles
are output on the fly at the time at which they cross the observer's lightcone.
For some simulations, lightcone outputs are available, such as the Hubble volume simulation \citep{Evrard2002}, the Euclid flagship simulation \citep{Potter2017}, and the DESI AbacusSummit 
simulations \citep{Maksimova2021,Hadzhiyska2022}, which are designed to meet the requirements of the new generation of large-scale structure surveys.
However, for many simulations, the positions of particles and haloes are
only output in the cubic box, at certain discrete time snapshots. In this case, the simulation
snapshots must be used to build approximate lightcones.
Creating approximate lightcones from the simulation snapshots also provides more flexibility, allowing the observer position to be chosen after the simulation has been run, e.g. in an analogue of the Local Group. Multiple lightcones can also be created with observers at different locations in the box. Creating lightcones on-the-fly is a very specialised use of a simulation, increasing the I/O and storage requirements, when simulation boxes are used for most applications. Additionally, lightcones currently available do not cover the full sky (e.g. the AbacusSummit lightcones cover 1 octant), so the snapshots can be used to create mocks that cover a larger footprint.

The simplest way to create a mock with galaxies positioned on the sky, cut to the survey geometry, is to use a single simulation snapshot, at the
median redshift of the galaxy sample being considered. In the final analysis of
the extended Baryon Oscillation Spectroscopic Survey \citep[eBOSS;][]{Dawson2016,eBOSS2021},
this is what was done to create mocks for the different galaxy tracers 
\citep{Smith2020,Alam2021,Rossi2021}. These mocks were used in the eBOSS mock challenges, to
validate the theoretical models used in the two-point clustering analyses. 
While this was good enough for the precision of eBOSS, these mocks lacked evolution
over the redshift ranges of the tracers. To include this evolution, multiple
snapshots can be used to build a lightcone.

There are two ways in which multiple snapshots can be combined to make a lightcone,
each with their own advantages and disadvantages. Firstly, haloes can be interpolated
between snapshots in order to build a lightcone \citep[e.g.][]{Merson2013,Smith2017,Izquierdo2019}. 
Interpolation has also been used to make LSST mocks in \citet{Korytov2019}.
This requires the use of halo merger trees to easily identify the descendants and progenitors
of a halo. However, halo interpolation is not perfect, and there are additional complications, 
such as halo mergers which take place between snapshots. Different methods for interpolating
haloes are compared in \citet{Smith2022}. Halo merger trees require a lot of computational effort to produce, and are not always available. Particle IDs are required to be able to track the same halo at different times. For some large simulations, halo IDs are not tracked in order to maximise the number of dark matter particles \citep[e.g.][]{Potter2017}, so merger trees cannot be produced.

The second method that can be used to build lightcones is to 
join together the snapshots in spherical shells. This is straightforward
to implement for any simulation, requiring less computational effort than interpolation, 
but has the disadvantage that there are discontinuities at the boundaries between shells. 
Galaxies or haloes that cross the lightcone close to a boundary can be duplicated, appearing at both sides, 
or conversely they can never appear at all \citep[e.g.][]{Kitzbichler2007}.

The method of joining snapshots in spherical shells has been applied to many simulations
to create lightcones and mocks for a wide range of applications. It is commonly
used to create lightcones for studies of weak lensing, e.g. in the `onion shell' MICE simulations
\citep{Fosalba2008,Fosalba2015}, and in the lensing lightcones of \citet{Giocoli2016,Giocoli2017}. 
A comparison of codes used to create lensing simulations can be found in \citet{Hilbert2020}. 
Lightcones have also been created in spherical shells in lensing studies of the cosmic microwave background 
\citep[e.g.][]{Carbone2008,Sgier2021}, and in studies of anisotropies in the gamma-ray background
\citep[e.g.][]{Zavala2010,Fornasa2013}. Other applications include mocks of active galactic nuclei \citep[e.g.][]{Comparat2019}
and lightcones of galaxy clusters \citep[e.g.][]{Zandanel2018}. Lightcones have been used in the BOSS survey 
to predict the HOD and clustering of the CMASS galaxies \citep{Rodriguez2016},
and have been created for the Dark Energy Survey \citep[DES;][]{Avila2018}. 
Mocks for the upcoming Roman Space Telescope have been made in \citet{Wang2022}.
This method is also being used to create galaxy lightcones for the Sloan Digital Sky Survey (SDSS) and the 
DESI Bright Galaxy Survey \citep{DongPaez2022}.

While the issue of repeated or missing galaxies in the lightcones has been known about for some time,
it is not accounted for in many of the lightcones that have been created.
One way to correct for this is to linearly interpolate the galaxies that are close to the interface \citep{Kitzbichler2007}.
In this paper, we quantify the effect of duplicated galaxies on the two-point clustering statistics of 
lightcone mocks that are constructed in spherical shells. 
We compare lightcones constructed using different numbers of 
snapshots, and assess the impact of duplicated galaxies on the small-scale clustering.
We propose a method to correct the issue of duplicated galaxies, which is simple to implement
for any simulation, and does not require halo merger trees and interpolation.

This paper is organised as follows. The simulations and lightcone construction is described
in Section~\ref{sec:mocks}. In Section~\ref{sec:comparing_mocks}, we compare lightcone 
mocks made with different numbers of snapshots, and assess the impact of duplicated galaxies
on the small-scale clustering measurements.
Our conclusions are summarised in Section~\ref{sec:conclusions}.

\section{Lightcones}
\label{sec:mocks}

In this paper, we create all-sky lightcones from the Millennium-XXL (MXXL) simulation.
Galaxies are added to the simulation using the HOD methodology of \citet{Smith2017} and
\citet{Smith2022}.

\subsection{MXXL simulation}

The MXXL simulation \citep{Angulo2012} is a dark-matter-only N-body simulation in a cubic
box of side length $3~\hGpc$ and particle mass $6.17\times10^9~\hMpc$.
The simulation was run in a WMAP1 cosmology \citep{Spergel2003}, 
with $\Omega_\mathrm{m}=0.25$, $\Omega_\Lambda=0.75$, $\sigma_8=0.9$, $h=0.73$ and $n_s=1$. 

Haloes in the simulation were first found using a friends-of-friends (FOF) algorithm
\citep{Davis1985}, with linking length $b=0.2$. Bound substructures within each FOF 
group were then identified using the \textsc{subfind} algorithm \citep{Springel2001}.

Halo catalogues are output at a total of 64 simulation
snapshots. There are 23 snapshots at $z<1$, which are 
approximately evenly spaced in expansion factor, $a$.

\subsection{Populating snapshots with galaxies}

The MXXL simulation snapshots are populated with galaxies using the HOD
methodology of \citet{Smith2017} and \citet{Smith2022}, 
where each galaxy is assigned an $r$-band magnitude. Here we briefly summarize this method.
A set of nested HODs for different absolute magnitude thresholds is used, which are 
measured from the SDSS survey \citep{Zehavi2011}. 
For a given magnitude threshold, the HOD is modelled as a smoothed step function 
for central galaxies, and a power law for satellites. 
The smoothing of the step function that represents the central galaxy HOD is performed using a pseudo-Gaussian spline kernel function rather than a simple Gaussian in order 
to prevent unphysical crossing of the HODs. 

Central galaxies are placed at the centre of a halo, while satellites are
randomly positioned following a NFW profile \citep{NFW1997}. The centrals are also assigned
the same velocity as the halo, with a random virial velocity for the satellites, relative to the central. This is drawn from a Gaussian distribution in each dimension, with the velocity dispersion of the halo. 

The HODs we use are evolved with redshift in order to match an evolving target
luminosity function, from measurements from the SDSS and GAMA surveys \citep{Blanton2003,Loveday2012}.
When applied to a single snapshot, the mock that is produced will match exactly the
target luminosity function at the redshift of the snapshot.

\subsection{Creating lightcones}

\begin{figure*} 
\centering
\includegraphics[width=0.48\linewidth]{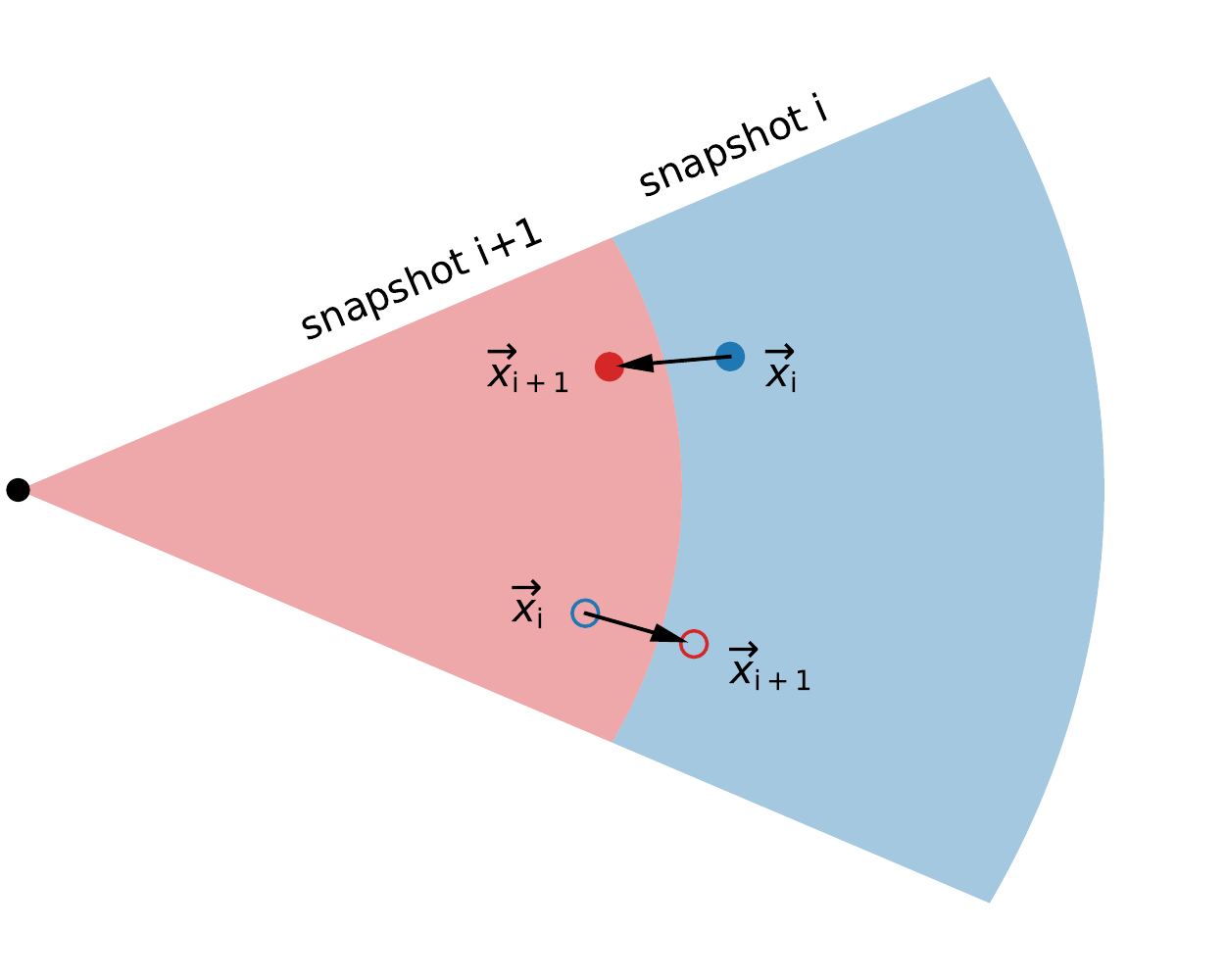}
\includegraphics[width=0.48\linewidth]{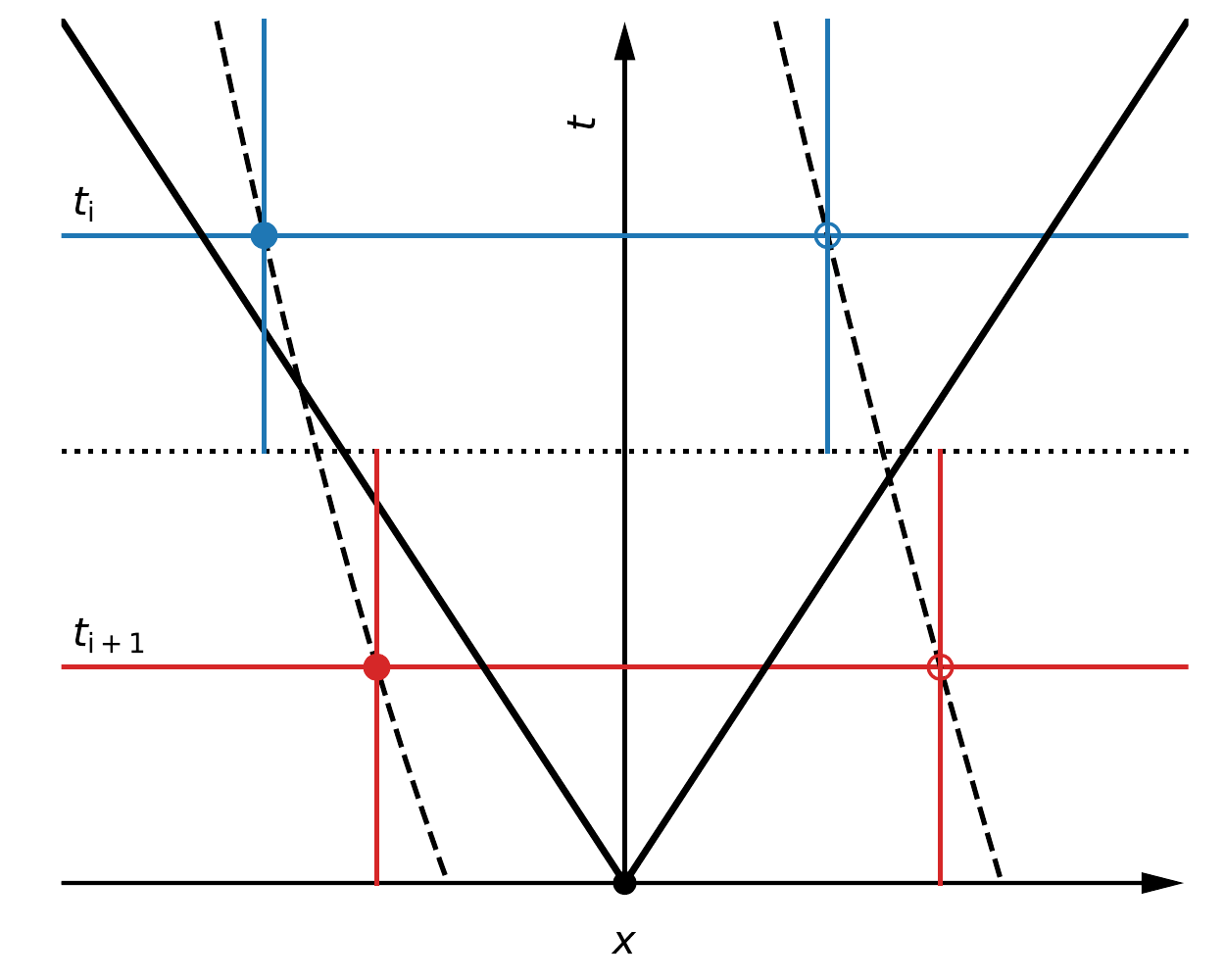}
\caption{\textit{Left}: Diagram depicting galaxy positions in a two-dimensional slice through a lightcone, illustrating how the same galaxy can appear twice, or never appear. The blue and red shaded regions are shells cut from two neighbouring snapshots, $i$ and $i+1$, respectively. The circles
show the positions of galaxies at the two snapshot output times, which 
cross the boundary.
The galaxy indicated by the filled circle moves towards the observer,
appearing in both shells. The galaxy indicated
by the open circle moves away from the observer, appearing in neither.
\textit{Right}: A depiction of the same two cases in a space-time diagram, where time is shown along the vertical axis, the red and blue horizontal lines are the
times corresponding the two simulation snapshots, and the diagonal lines represent the observer's lightcone. 
Within each snapshot, the position of each galaxy is kept fixed (vertical coloured lines), with an instantaneous jump between snapshots. 
Consequently, the galaxy depicted by the solid symbol, which is moving towards the observer, crosses the observer's past lightcone twice, while the one depicted by the open symbol does not cross it at all.
If the galaxy trajectories had instead been interpolated (dashed lines) they would have each crossed the lightcone once. In this figure, the galaxy velocities have been greatly exaggerated.
In reality, the dashed lines would be close to
vertical and only galaxies very close to the lightcone at the interface between shells would be subject to these problems.
}
\label{fig:lightcone}
\end{figure*}

After populating the snapshots with galaxies, we create 4 full-sky lightcones by joining 
together the snapshots in spherical shells. This is done using 1, 3, 5 and 9 snapshots,
where the snapshots used in each lightcone are summarised in Table~\ref{tab:redshifts}.
The joins between snapshots occur at the redshift exactly half way between the snapshot
redshifts. For the lightcone created using a single snapshot, we use the snapshot which is the closest
to the median redshift of the volume limited galaxy sample we examine in this work (see Section~\ref{sec:galaxy_sample}).
Since the MXXL simulation is large, no periodic replications of the box are required.

When cutting each snapshot into spherical shells, there is a choice of whether this is
done based on the real-space position of each galaxy, or based on the positions in 
redshift space. We therefore create two versions of each lightcone, to assess the impact
of this choice on the galaxy clustering statistics. 

When the lightcone is constructed from a single snapshot, the galaxy luminosity function is 
constant with redshift. For the other lightcones, there are sudden jumps in the luminosity
function at the boundaries between snapshots. To make sure that all mocks have the same
luminosity function, which smoothly evolves with redshift, we apply an abundance matching rescaling to the
absolute magnitudes, as in \citet{DongPaez2022}. The magnitude of a galaxy in a shell, $M_r$,
can be converted to a corresponding cumulative number density, $n$, using the target luminosity function
at the snapshot redshift, $z_\mathrm{snap}$. The 
target luminosity function at the redshift of the galaxy in the lightcone, $z$, is then used to
convert the cumulative number density back a magnitude at redshift $z$.

\begin{table}
\centering
\begin{tabular}{ccccc}
\hline
Redshift & 1 Snapshot & 3 Snapshots & 5 Snapshots & 9 Snapshots \\
\hline
$0$      & & &\checkmark & \checkmark \\
$0.0199$ & &\checkmark & & \checkmark\\
$0.0414$ & & &\checkmark & \checkmark\\
$0.0644$ & & & & \checkmark\\
$0.0892$ & &\checkmark &\checkmark & \checkmark\\
$0.1159$ & & & & \checkmark\\
$0.1444$ &\checkmark & &\checkmark & \checkmark\\
$0.1749$ & &\checkmark & & \checkmark\\
$0.2075$ & & & \checkmark & \checkmark\\
\hline
\end{tabular}
\caption{MXXL simulation snapshots used in the construction of the lightcone mocks,
built with a total of 1, 3, 5 and 9 snapshots. The redshift of the 1-snapshot mock
is closest to the median redshift of the galaxy sample we use.}
\label{tab:redshifts}
\end{table}

\subsection{Duplicated galaxies}

In the lightcones built from multiple snapshots, it is possible
for some galaxies to appear twice in the lightcone, or to never appear at all. 
This happens when a galaxy crosses the interface between two shells, as illustrated in Fig.~\ref{fig:lightcone}.
In the diagram on the left, the blue and red shaded regions indicate two shells in the lightcone, which
are cut from neighbouring snapshots $i$ and $i+1$, at output times $t_i$ and $t_{i+1}$.
The points show the positions of galaxies at these two times. 
For the galaxy shown by the solid points, its position vector in the initial snapshot, $\overrightarrow{x}_i$, falls within the shell cut from snapshot $i$, so the galaxy appears in the lightcone.
At time $t_{i+1}$, the galaxy has moved towards the observer, and its new position vector, $\overrightarrow{x}_{i+1}$, is inside the shell from snapshot $i+1$. This galaxy appears
twice in the lightcone, at each side of the boundary between shells.
The opposite happens for the galaxy shown by the open circles. Its position at $t_i$ falls
outside of the first shell. The galaxy then moves away from the observer, and its new position at 
$t_{i+1}$ is also outside of the second shell. This galaxy never appears in the lightcone. In reality, both galaxies should appear once in the lightcone. We can see from this figure that galaxies are only duplicated if they cross the boundary while travelling towards the observer, and missing galaxies always travel away from the observer. For the galaxies that appear twice, their pair separation is simply the distance travelled between the two snapshots. 

The space-time diagram on the right of Fig.~\ref{fig:lightcone} illustrates the
same two cases, with position on the $x$-axis and time on the $y$-axis.
The galaxy that is moving towards the observer and appears twice in the lightcone is shown by the solid circles. Within each snapshot, the position of the galaxy is kept constant, which is indicated by the vertical lines. A galaxy in the 
shell from snapshot $i$ will appear in the lightcone at a time between
$(t_{i-1} + t_i) / 2 < t < (t_i + t_{i+1}) / 2$, but its position is fixed at 
$\overrightarrow{x}_i$. At the boundary, the galaxy jumps instantaneously to its 
new position, $\overrightarrow{x}_{i+1}$, travelling faster than the speed of light.
In this example, the galaxy crosses the lightcone in snapshot $i$, then
jumps back over the lightcone, crossing a second time in snapshot $i+1$.
If the galaxy was interpolated, it would follow a smooth trajectory, indicated
by the dashed line, crossing the lightcone once. Similarly, the open circles
show the galaxy that never appears, since it instantaneously jumps over the
lightcone at the interface where the shells are joined together.

\subsection{Galaxy sample}
\label{sec:galaxy_sample}

We cut the mock to a volume limited sample of $z<0.2$ and $M_r < -20$,
where the number density of galaxies is constant with redshift.
The wide redshift range is covered by a total of 9 snapshots,
allowing us to investigate the effect of making lightcones
from different numbers of shells.
We also cut to central galaxies only, and do not consider the satellites.
On small scales, most galaxy pairs are from satellite galaxies within the same 
halo, which will reduce the effect of duplicated galaxies on the clustering
measurements. We focus on the central galaxies, where these effects will be
strongest.

In this paper we focus on a volume limited galaxy sample, but we have checked
that our results and conclusions remain unchanged for an apparent magnitude threshold galaxy sample.

\section{Comparing lightcones with different numbers of snapshots}
\label{sec:comparing_mocks}

In this section we compare the lightcones constructed from different numbers of snapshots. 
Section~\ref{sec:clustering} compares the two-point clustering statistics, where
the snapshots are carved into shells based on the real-space or redshift-space
galaxy positions. Section~\ref{sec:duplicated_galaxies} quantifies the distances
that galaxies travel between snapshots, which sets the scales at which the clustering
measurements are affected by duplicated galaxies. We assess the impact of 
removing duplicated galaxies on the two-point clustering statistics 
in Section~\ref{sec:removing_duplicates}.

\subsection{Galaxy clustering}
\label{sec:clustering}

\begin{figure} 
\centering
\includegraphics[width=\linewidth]{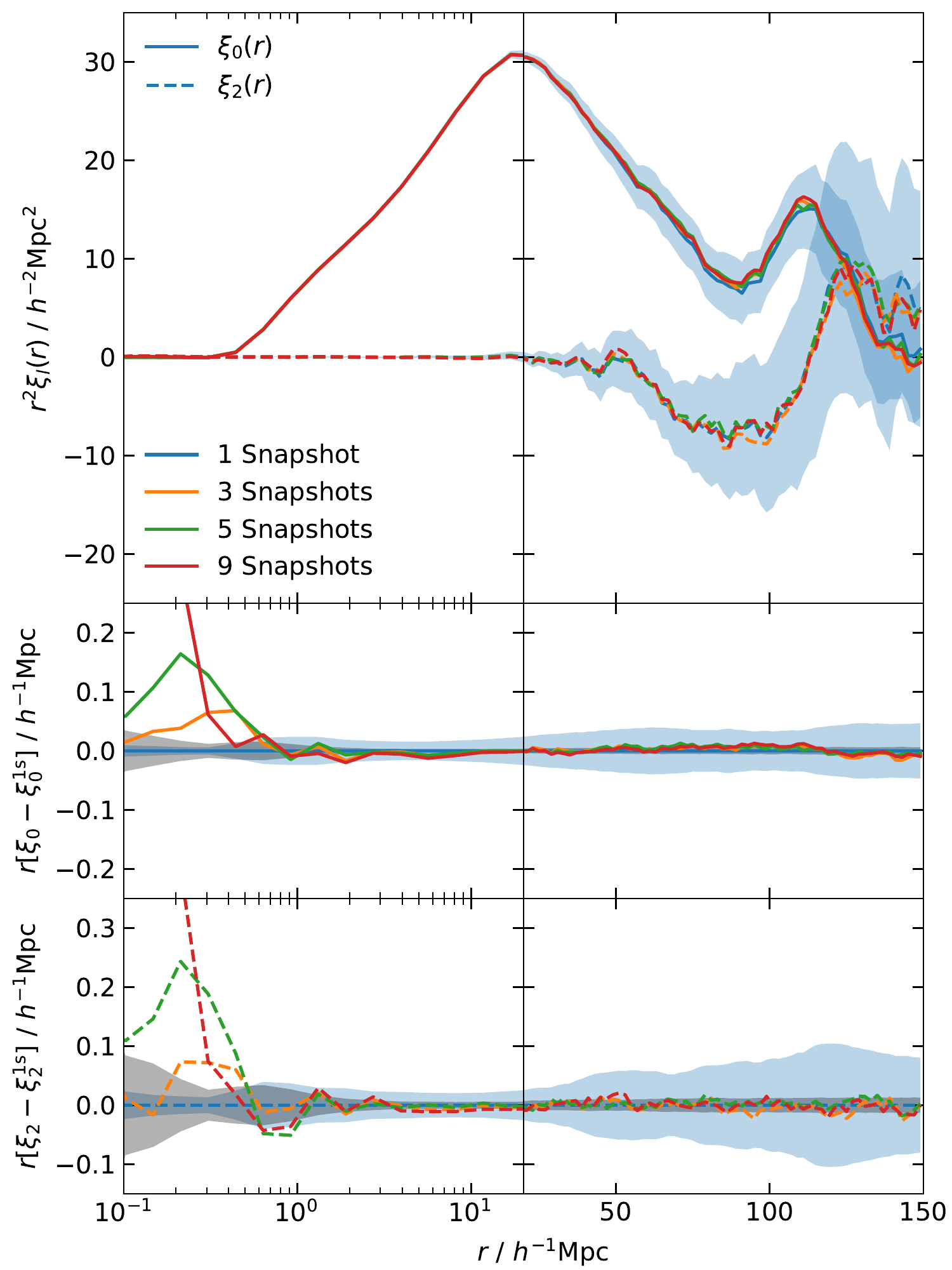}
\caption{Correlation function multipoles in real space, scaled by $r^2$, for the lightcones built from
1 snapshot (blue), 3 snapshots (orange), 5 snapshots (green) and 9 snapshots (red).
The snapshots are cut into shells based on the galaxy positions in real space.
The upper panel
shows the monopole (solid lines) and quadrupole (dashed), transitioning from
a logarithmic scale to a linear scale on the $x$-axis at $20~\hMpc$. The blue shaded region is
the jackknife error, calculated using 100 jackknife samples. 
The grey shaded region indicates the jackknife error in $r\Delta \xi$, averaged over all pairs of mocks.
The middle and lower panels show the
differences in the monopole and quadrupole, respectively, compared to the mock built from a single
snapshot, and scaled by $r$. }
\label{fig:xi_shells_real}
\end{figure}

We measure the correlation function, $\xi(s,\mu)$, from the lightcones, where $\mu$ is the cosine 
of the angle between the line-of-sight direction, and pair separation vector. This is then decomposed into 
Legendre multipoles,
\begin{equation}
\xi_l(s) = \frac{2l+1}{2} \int_{-1}^1 \xi(s,\mu) P_l(\mu)d\mu,
\end{equation}
where $P_l(\mu)$ is the $l$\textsuperscript{th} order Legendre polynomial. In redshift space, the first two even multipoles are
the monopole, $\xi_0(s)$, and quadrupole, $\xi_2(s)$, which are non-zero in linear theory. In
real space, the monopole is equivalent to the real-space correlation function, $\xi(r)$, while the quadrupole
is on average zero.

\begin{figure*} 
\centering
\includegraphics[width=0.48\linewidth]{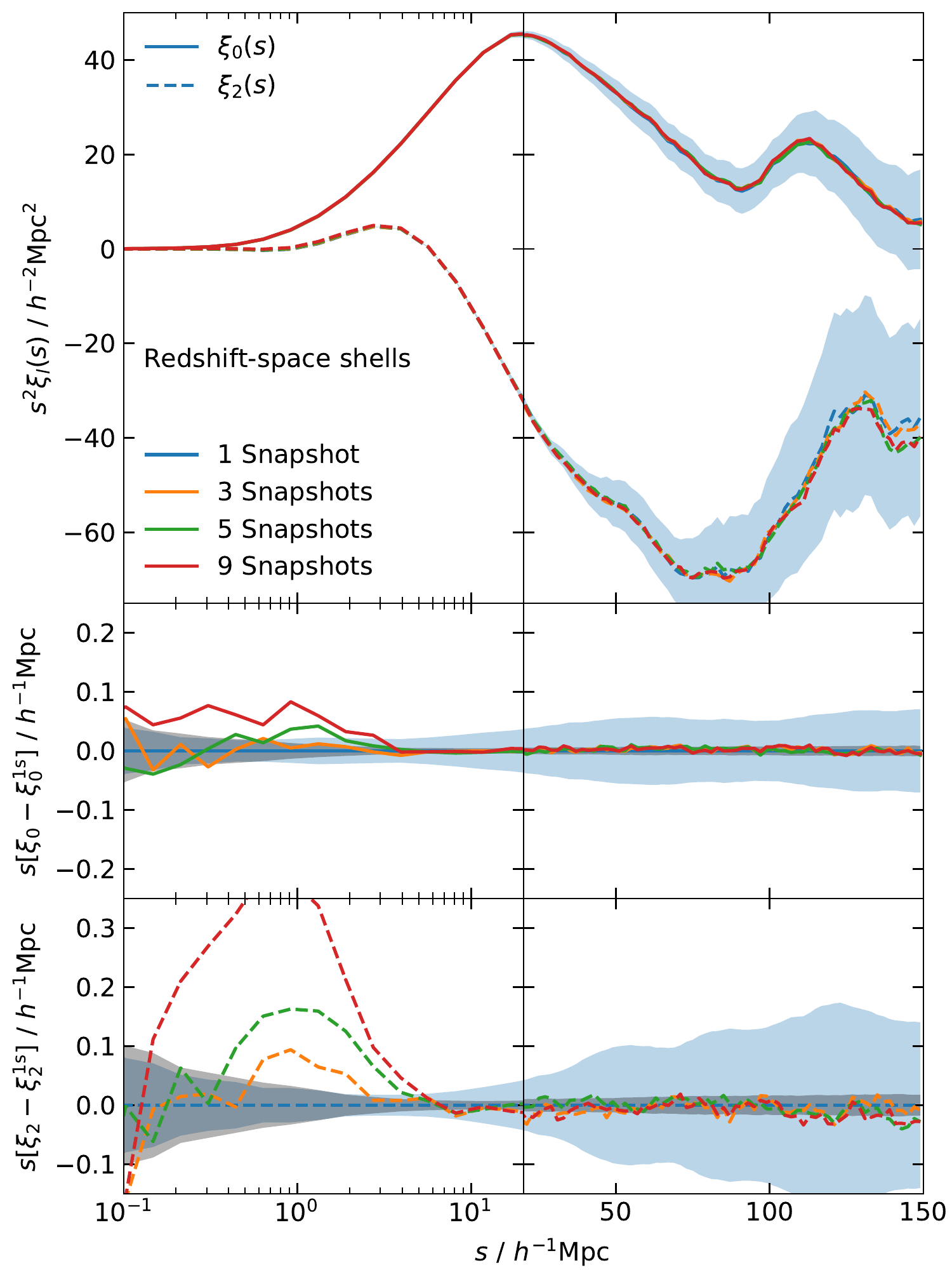}
\includegraphics[width=0.48\linewidth]{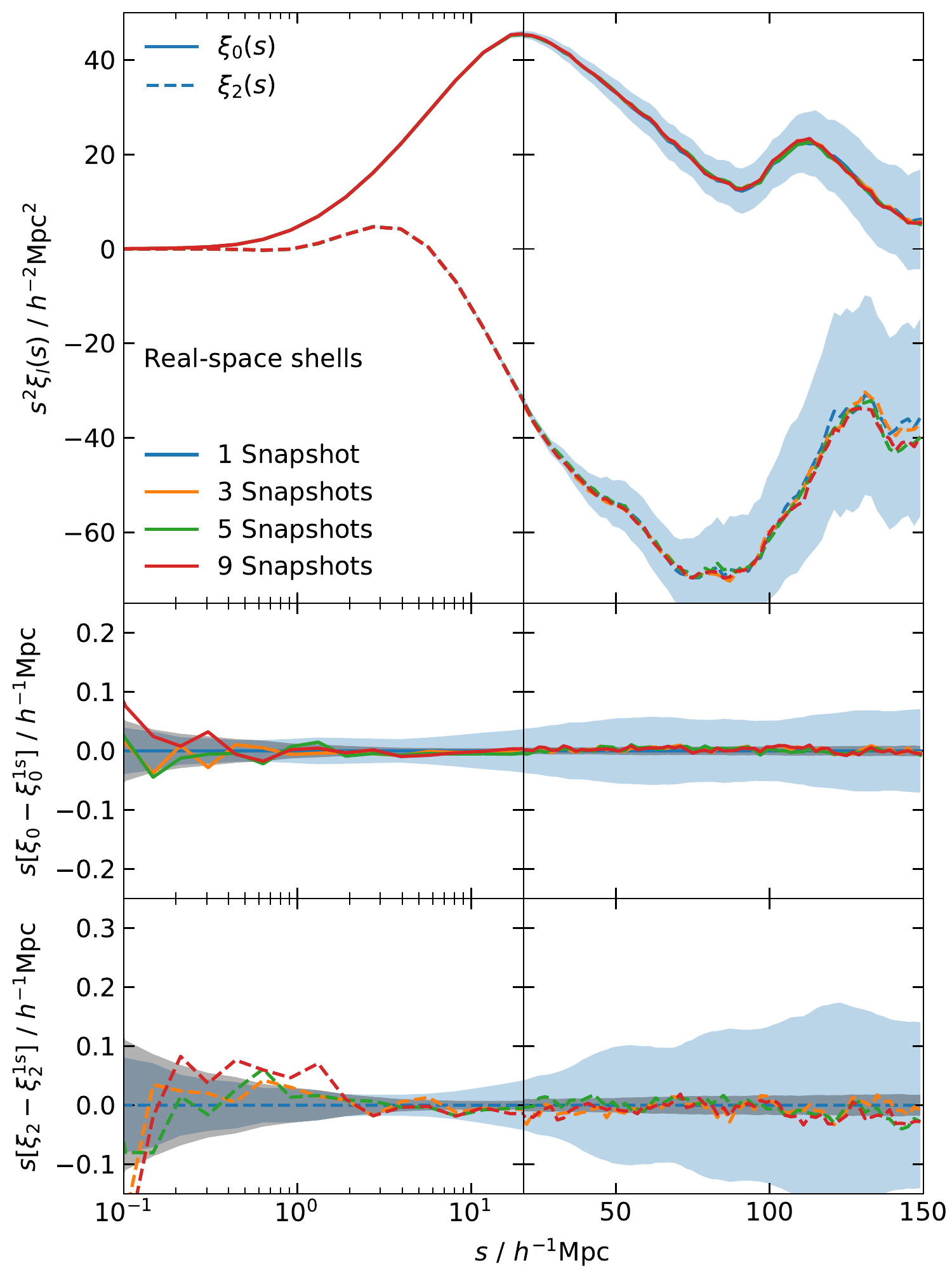}
\caption{As Fig.~\ref{fig:xi_shells_real}, showing the correlation function 
multipoles in redshift space. The lightcones are constructed
by cutting the snapshots in shells in redshift space (left) and in real space (right).}
\label{fig:xi_shells_redshift}
\end{figure*}

Fig.~\ref{fig:xi_shells_real} shows the two-point correlation function 
in real space of the 4 lightcones, where the snapshots are cut into shells based
on the real-space position of each galaxy.
The monopole and quadrupole are shown in the 
upper panel, where the blue shaded region indicates the
jackknife error from 100 jackknife regions. In real space, the quadrupole 
should be zero, and the signal measured in the mocks comes from cosmic variance
in the finite volume. 
Since all 4 mocks are constructed from the same simulation, with the observer 
positioned at the same location, they all share the same large scale structure,
so the correlation function measurements have the same shape.
The middle panel shows the difference in $\xi_0(r)$, relative to
the mock built from a single snapshot, which is scaled by a factor of $r$ to highlight any differences on large scales.
We use the 1-snapshot lightcone as the reference and
differences from this are artifacts due to duplicated galaxies.
Since the snapshot used is at the median redshift of the galaxy sample,
the clustering measurements are a good approximation of a true lightcone.
The grey shaded region is the 
jackknife error in $r\Delta \xi_0$, which provides an estimate of the noise
when comparing simulations with the same initial conditions
\citep[see equation 14 of][]{Grove2021}. This error is calculated for all
pairs of lightcones, and the average is plotted.
On large scales, this noise is much smaller than the uncertainty due to 
cosmic variance, which is estimated with the standard jackknife error.
On large scales,
all 4 lightcones show good agreement in the monopole, and they remain in
good agreement down to $\sim 1~\hMpc$. Below this, the mocks built from multiple 
snapshots peel off, since their clustering is boosted by pairs of galaxies which
are duplicated at the interfaces between shells. The scale at which this 
occurs is the smallest for the mock with 9 snapshots, but for this lightcone,
the effect on the monopole is also the strongest.
The bottom panel shows the differences in $\xi_2(r)$. As with the monopole,
all the mocks are in good agreement on large scales, but there is a 
non-zero signal below $1~\hMpc$ for the lightcones built from multiple 
snapshots, which again is the strongest and pushed to the smallest scales
for the lightcone that uses all 9 snapshots. 
The increase in the quadrupole indicates that the pair separation of duplicated galaxies is preferentially directed along the line of sight. This is because galaxies that travel along
the observer's line of sight are more likely to cross a boundary
between snapshots, and appear twice in the lightcone.

The left panel of Fig.~\ref{fig:xi_shells_redshift} shows the clustering in redshift 
space. Here, the snapshots were cut into shells using their redshift-space
positions. Including the effect of velocities smooths out the effect of 
duplicated galaxies on the monopole. The clustering is still boosted for the
lightcone with 9 snapshots, but the increase in the clustering is smaller than
in real space. However, this difference extends to larger scales, around
$4~\hMpc$. A large effect is still seen in the quadrupole, which is also
shifted to larger scales compared to in real space. A clear monotonic trend can be seen, where the 
strength of the quadrupole increases as more snapshots are included.
Above $\sim 10~\hMpc$, all the lightcones remain in good agreement.

When making the lightcones in redshift space, the shells were previously cut
based on the redshift-space position of the galaxies in each snapshot. 
Alternatively, the cuts can be done based on the positions in real space,
with the effect of velocities added to the mocks afterwards. The 
redshift-space clustering is shown in the right panel of Fig.~\ref{fig:xi_shells_redshift} for the 
lightcones where the cuts were applied in real space. The effect of applying
the velocities at the end greatly reduces the effect that duplicated galaxies have on
the clustering. For the monopole, the clustering measurements for all
the lightcones are in good agreement down to very small scales of $\sim 0.2~\hMpc$.
Below this, a small boost in the clustering can only be seen for the mock
made of 9 snapshots. There is still an excess in the quadrupole at scales
of $\sim 1~\hMpc$, which is strongest for the 9-snapshot lightcone, but
this is much smaller than when the snapshots were joined in redshift space.

\subsection{Distance separation of duplicated galaxies}
\label{sec:duplicated_galaxies}

On small scales, the two-point clustering statistics are boosted due to galaxies
that appear twice in the lightcone, at each side of the interface where
two snapshots are joined together. The separation between the galaxies
in each duplicated pair is simply the distance that the galaxy travelled in the time between
the two snapshots. Therefore, the distribution of these distances will provide
insight into how the clustering measurements are affected by the 
number of snapshots used to build the lightcones.

\begin{figure} 
\centering
\includegraphics[width=\linewidth]{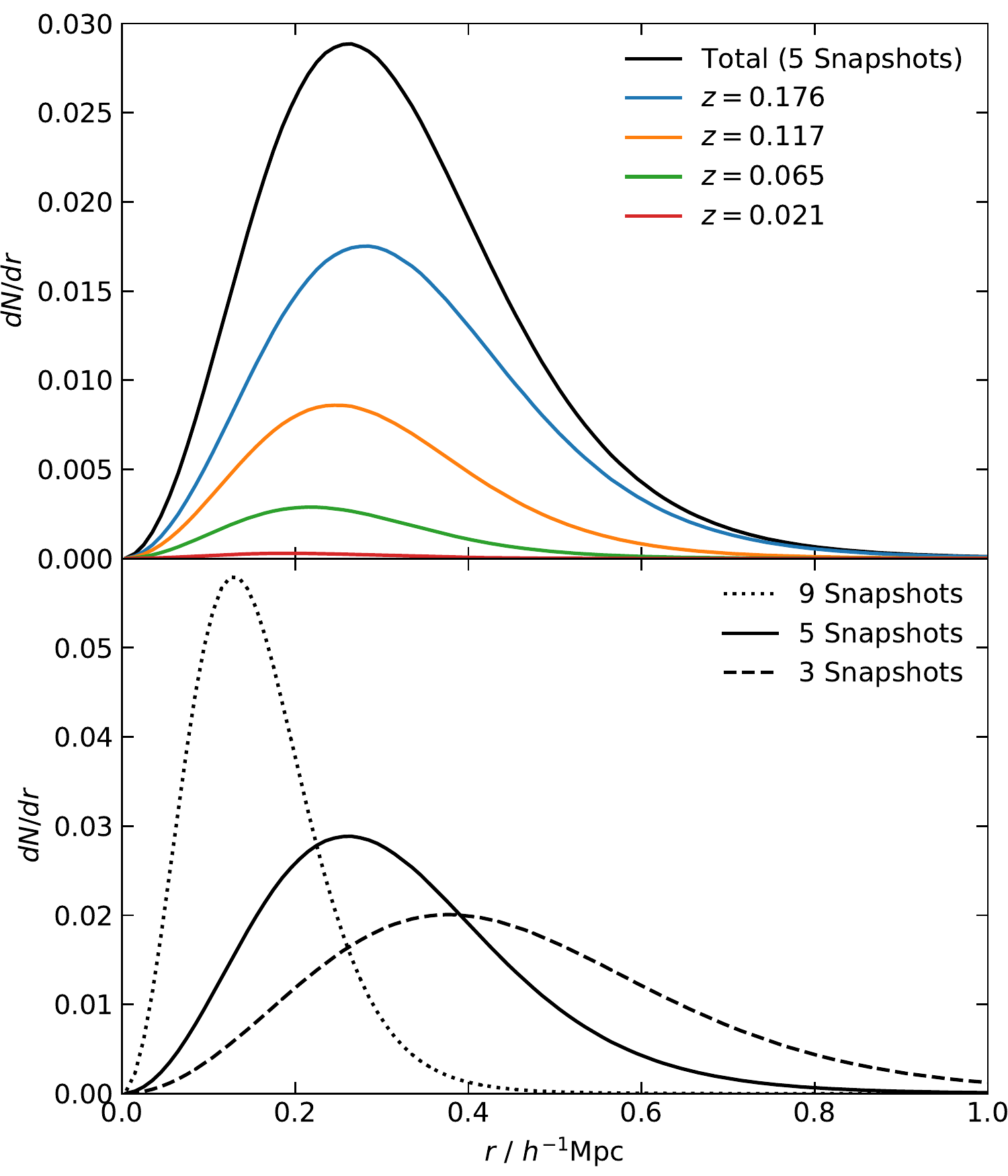}
\caption{\textit{Top}: Normalised distribution of the distances, $r$, travelled by central galaxies at the boundaries
between shells, for the lightcone constructed from 5 snapshots. The coloured curves show
the distributions at the 4 boundaries, as indicated in the legend. The black curve is the
total distribution (the sum of the coloured curves).
The coloured curves are normalised so that the area under the
black curve is 1.
\textit{Bottom}: Total distributions (sum of the coloured curves in the top panel) for mocks built from 9 snapshots (dotted curve),
5 snapshots (solid curve, which is the same as in the upper panel) and 3 snapshots (dashed curve).}
\label{fig:distances}
\end{figure}

Fig.~\ref{fig:distances} shows the normalised distribution of distances that a central galaxy
(i.e. a halo) travels between each of the simulation snapshots used to make
the lightcone. This distribution is calculated from the full
snapshots, using all central galaxies brighter than the magnitude threshold
of $M_r<-20$, allowing the distributions to be measured smoothly. 
The distance that each galaxy travels is calculated by multiplying
its velocity by the time interval, $\Delta t$, between the adjacent snapshots
used to build the lightcone. The amplitude of the distribution is then scaled by a factor of $r_\mathrm{com}^2$,
where $r_\mathrm{com}$ is the comoving distance from the observer to the boundary in the lightcone,
to take into account the differences in area (boundaries at high redshift cover
a larger area, and hence there will be more galaxies that are duplicated). 

The upper panel of Fig.~\ref{fig:distances} shows the distribution of the distances that galaxies travel
between pairs of snapshots (i.e. the separations between duplicated
galaxies) for the 5-snapshot lightcone. The coloured curves show this distribution for each consecutive
pair of snapshots, where the redshift at each boundary in the lightcone
is indicated in the legend. The black curve is the total distribution 
(the sum of these). All the curves have been normalised to that the area under the sum (the black curve) is 1.

For the 5-snapshot lightcone, the total distribution peaks at $\sim 0.25~\hMpc$. 
Most duplicated pairs come from the highest redshift interface,
at $z=0.176$, since it covers a larger area in the lightcone.
The number of duplicates is smaller at low redshifts, and the
peak of the distribution also shifts to smaller scales. This is because
the simulation snapshots are not evenly spaced, while the streaming velocities of haloes only evolve weakly with redshift. At low redshifts, the
snapshots are spaced closer together, so there is less time 
in which the galaxies are able to travel. 

The lower panel of Fig.~\ref{fig:distances} shows the total distributions (the sum of the coloured curves in the upper panel),
comparing the lightcones that were made using 3, 5 and 9 snapshots.
When only 3 snapshots are used, the total number of duplicated
galaxies is small, but the distribution peaks at $\sim 0.4~\hMpc$,
with a long tail extending to to $1~\hMpc$. As the number of snapshots
used is increased, the peak of the distribution shifts to smaller
scales, since the $\Delta t$ between snapshots at each boundary is smaller.
When all 9 snapshots are used, the peak shifts down to $\sim 0.15~\hMpc$,
with almost no pairs with separation above $0.5~\hMpc$.
The total number of duplicated galaxies stays approximately
constant when different numbers of snapshots are used. 
This is because if the number of interfaces is doubled,
half as many galaxies cross at each one (since the
$\Delta t$ between snapshots halves, so each galaxy 
travels half the distance).

These observations are consistent with the clustering measurements in real
space in Fig.~\ref{fig:xi_shells_real}. 
The scale at which the distributions peak 
($\sim 0.4~\hMpc$ and $\sim 0.25~\hMpc$ for 3 and 5 snapshots, respectively),
is also where we see the largest difference in the monopole, compared to 
the single-snapshot lightcone. For the 9-snapshot lightcone, most duplicated
pairs have separation less than $\sim 0.4~\hMpc$, which is the same scale
where the monopole peels upwards.

\subsection{Removing duplicated galaxies}
\label{sec:removing_duplicates}

\begin{figure} 
\centering
\includegraphics[width=\linewidth]{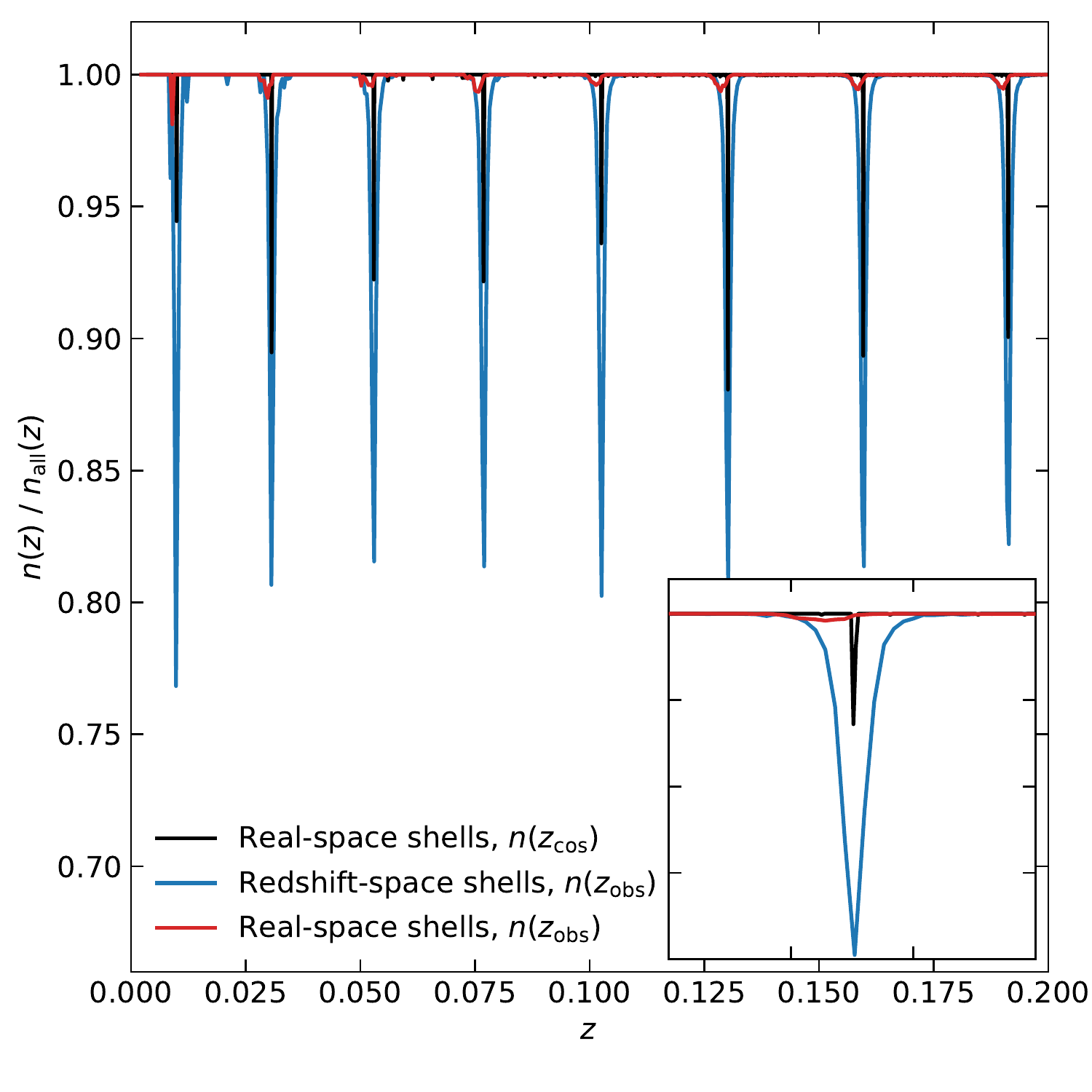}
\caption{Ratio of the $n(z)$ of the galaxy sample, before and after the removal of duplicated galaxies, showing the fraction of galaxies that are removed. The black line is the ratio of the $n(z_\mathrm{cos})$ (i.e. in real space) from the lightcone with shells cut in real space. The blue and red lines show the ratios of the $n(z_\mathrm{obs})$ (i.e. in redshift space), from the lightcones with shells cut in redshift and real space, respectively.
The smaller panel shows the dip in the $n(z)$ ratios at $z=0.1$,
but zoomed in on the $x$-axis.}
\label{fig:nz_ratio}
\end{figure}

\begin{figure*} 
\centering
\includegraphics[width=0.45\linewidth]{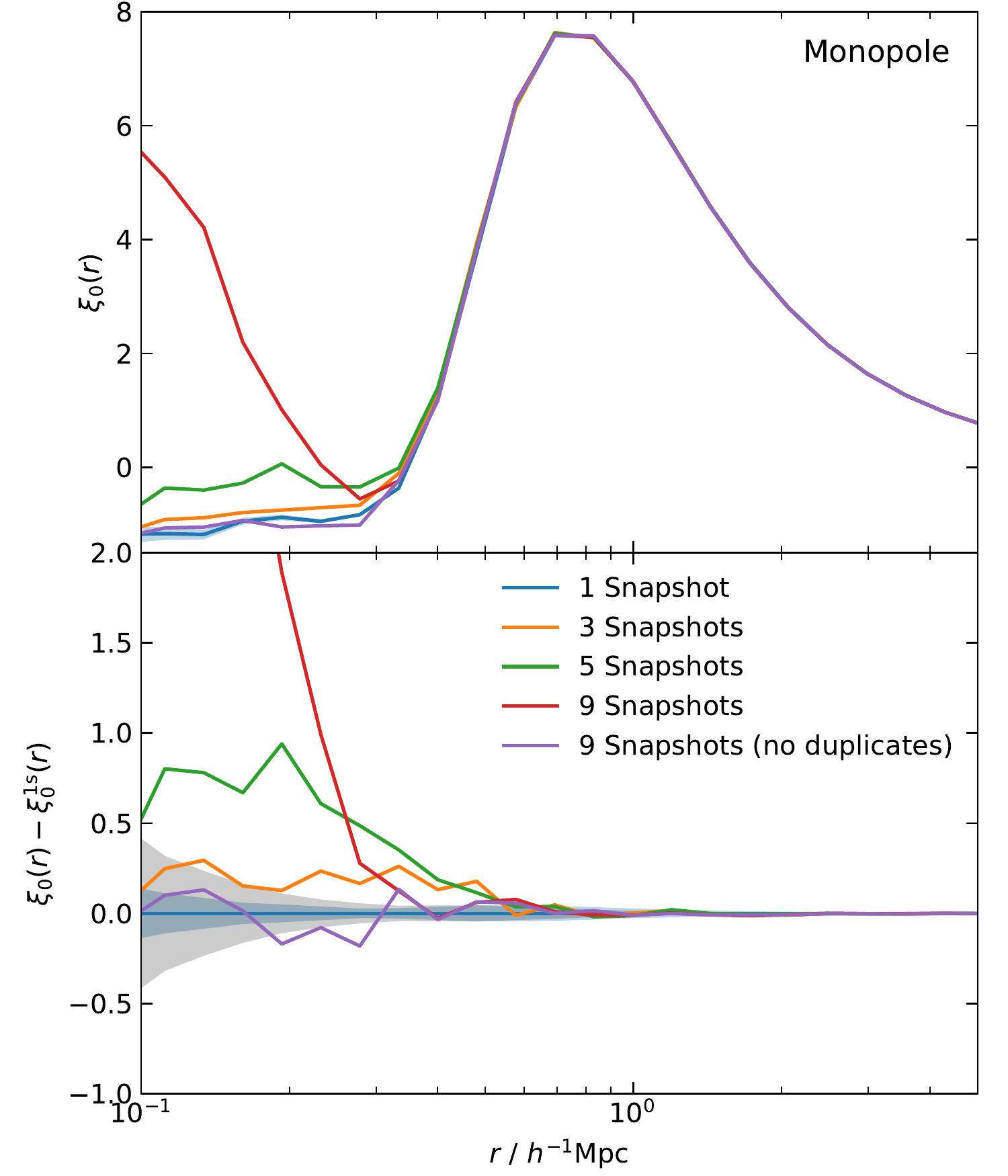}
\includegraphics[width=0.45\linewidth]{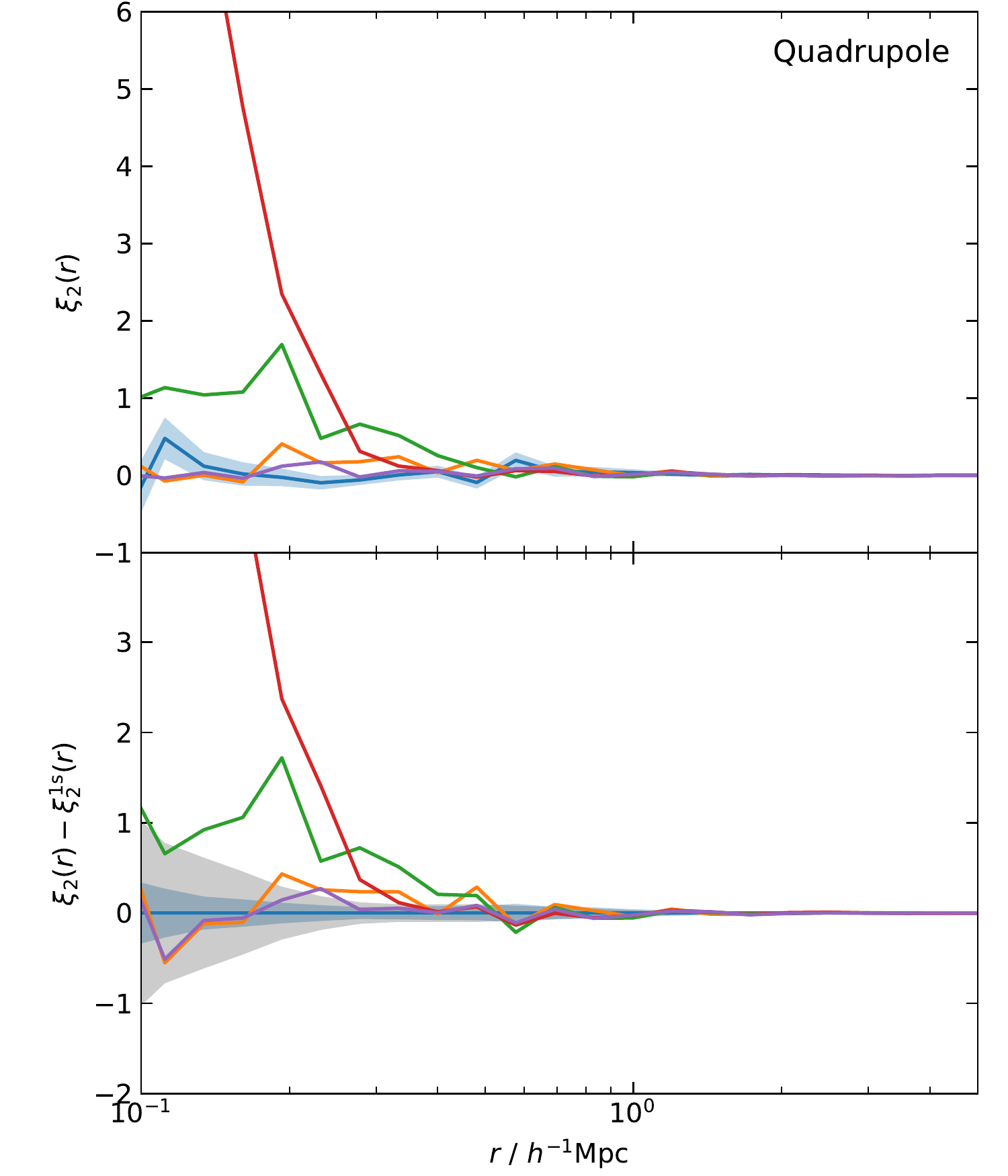}
\caption{\textit{Left}: Real-space monopole on small scales, for mocks built from 1 snapshot (blue),
3 snapshots (orange), 5 snapshots (green), 9 snapshots (red), and 9 snapshots with duplicated
galaxies removed (purple). The upper panel shows the monopole, while the lower panel is the difference
relative to the mock built from a single snapshot. The blue shaded region in the jackknife error, using
100 jackknife regions, and the grey shaded region is the jackknife error in $\Delta \xi$.
\textit{Right}: Like the left plot, but showing the correlation function quadrupole.}
\label{fig:xi_small_real}
\end{figure*}

\begin{figure*} 
\centering
\includegraphics[width=0.45\linewidth]{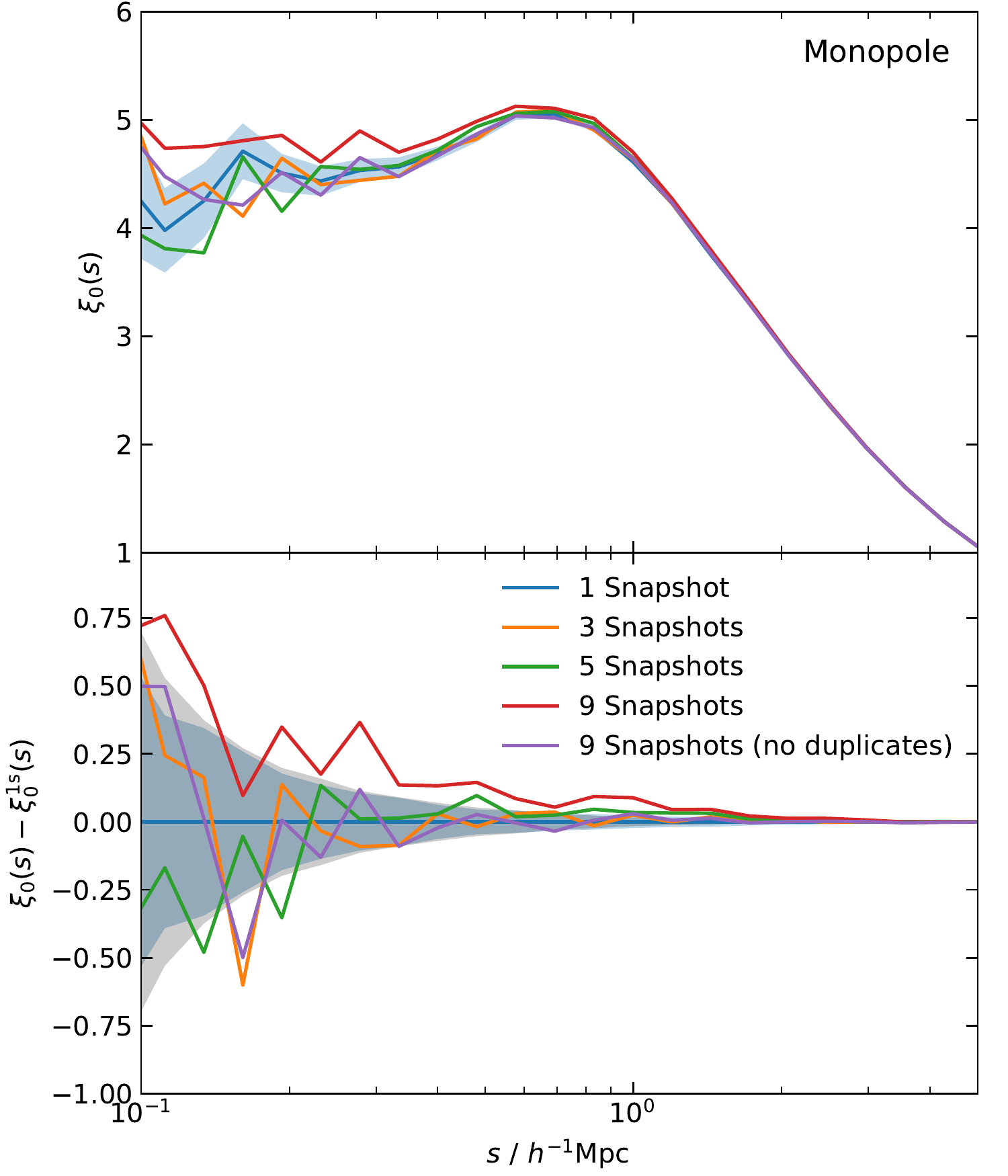}
\includegraphics[width=0.45\linewidth]{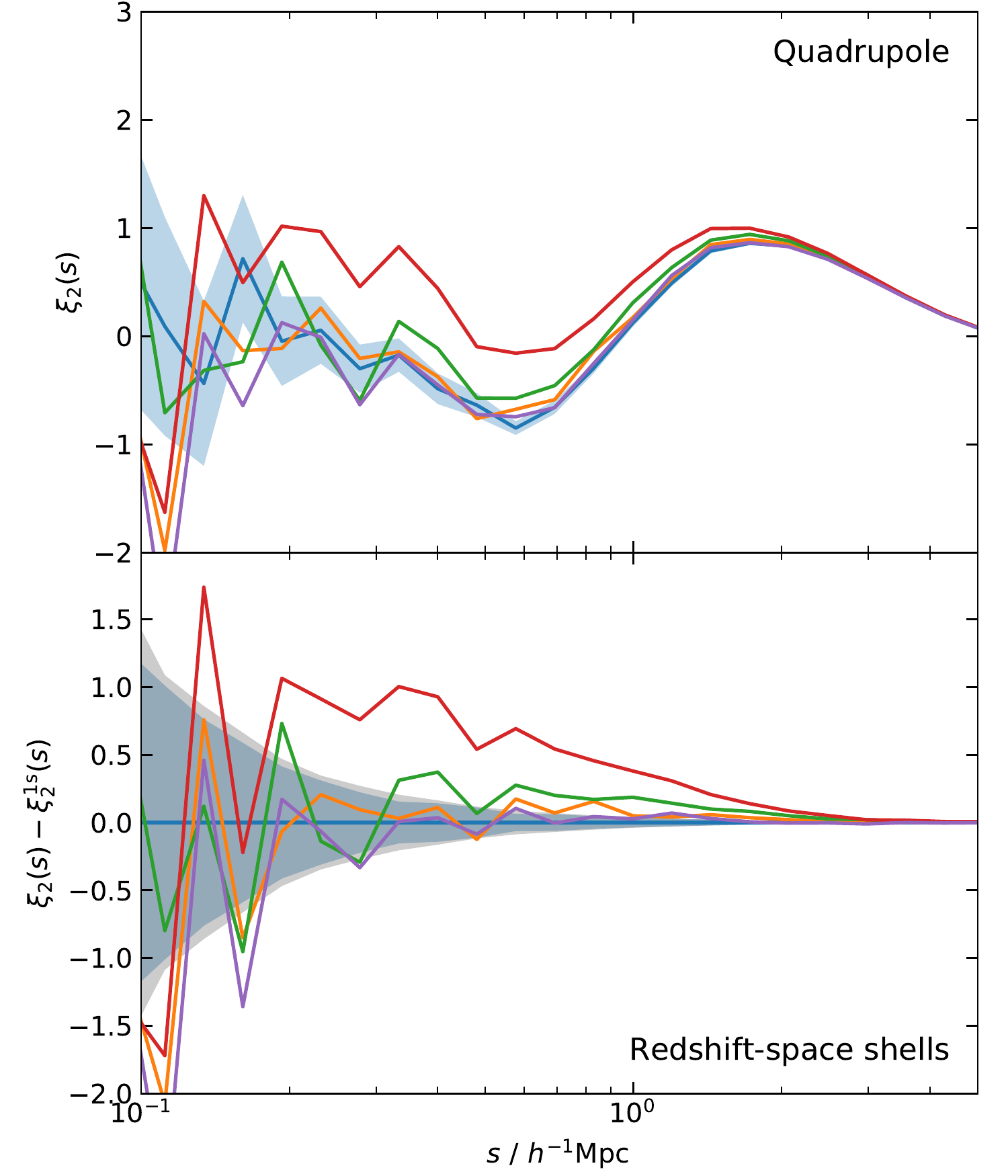}\\
\vspace{0.1cm}
\includegraphics[width=0.45\linewidth]{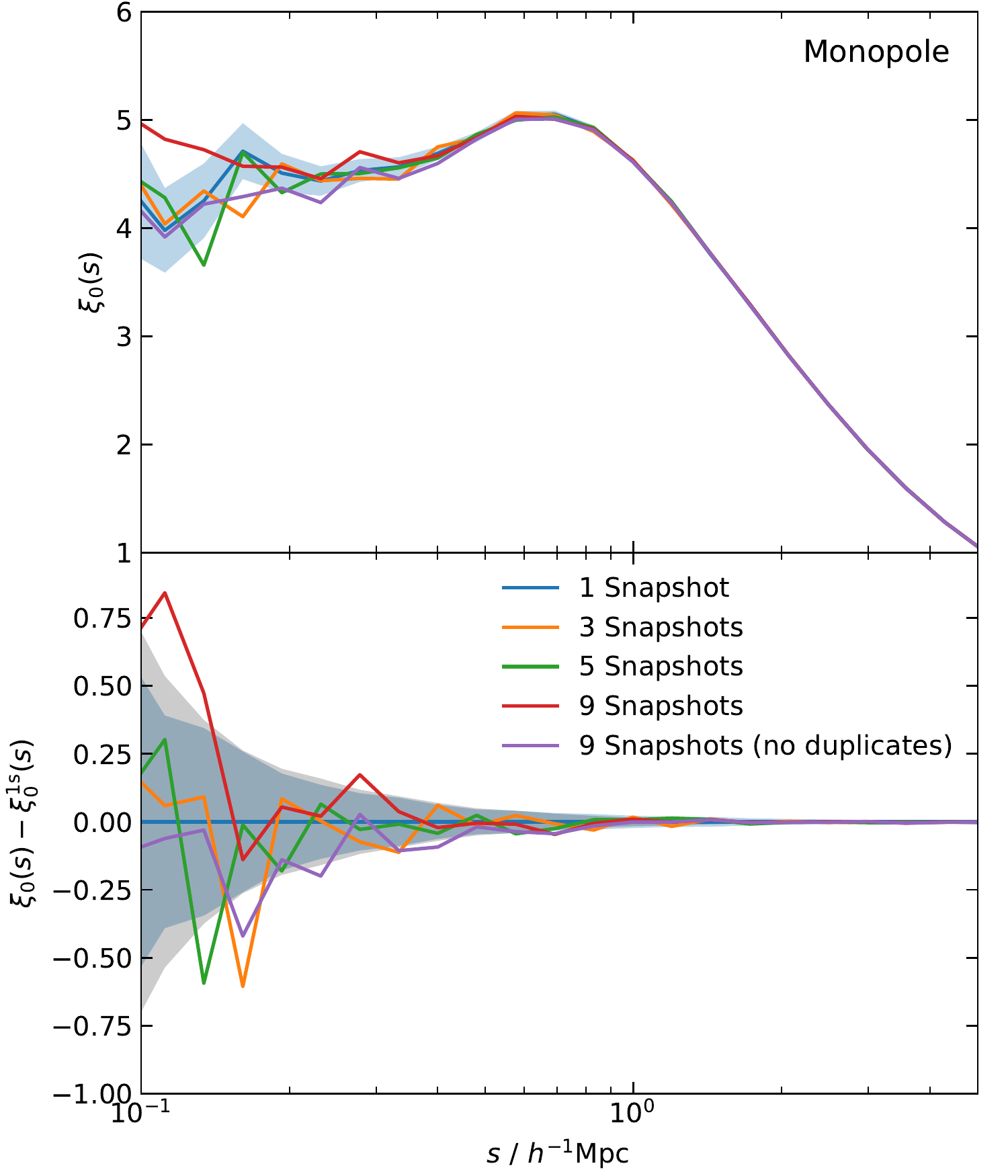}
\includegraphics[width=0.45\linewidth]{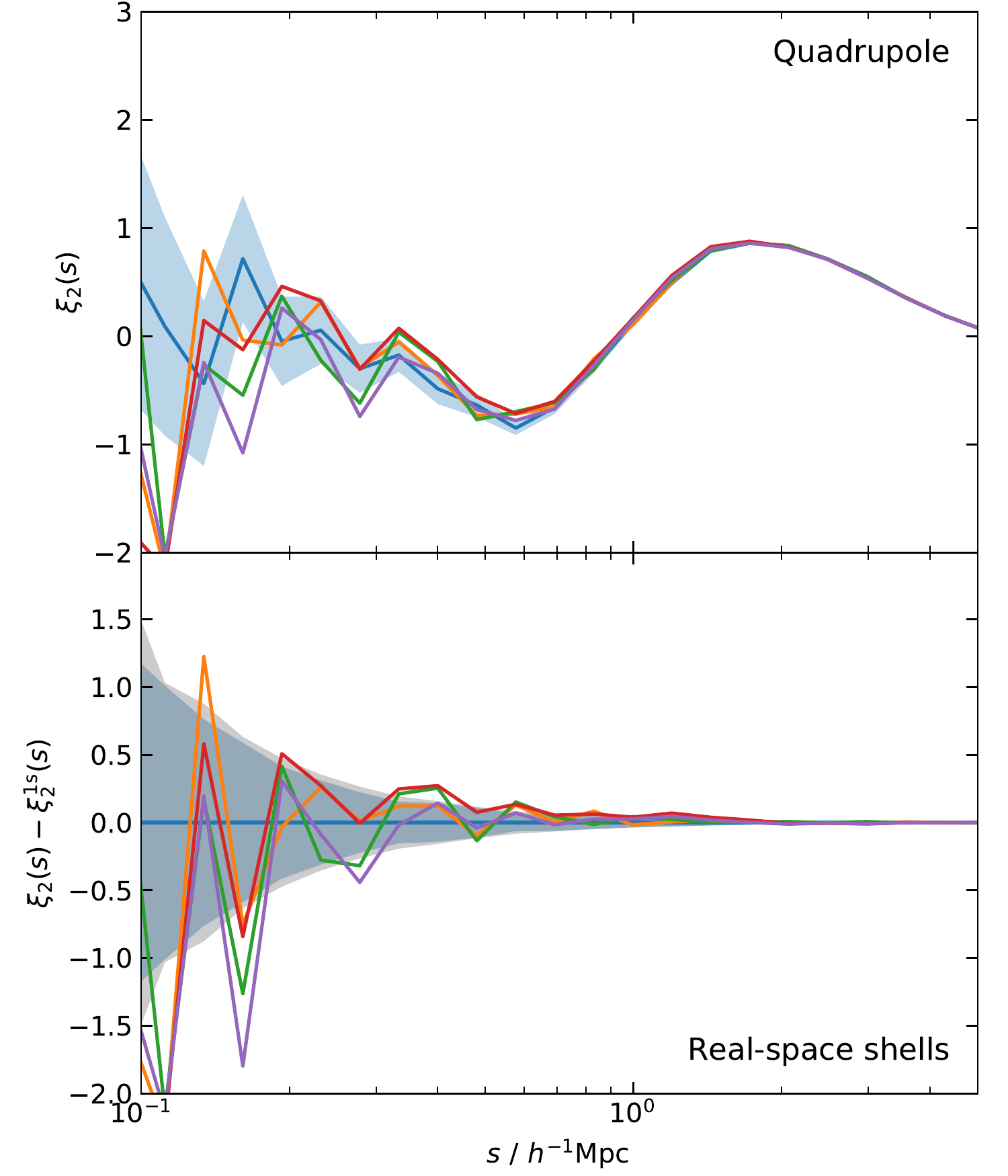}
\caption{As Fig.~\ref{fig:xi_small_real}, but showing the small-scale clustering in
redshift space, where the spherical shells were joined
in redshift space (top panels), and where the shells were joined in real space (bottom panels)}
\label{fig:xi_small_redshift}
\end{figure*}

We now test the effect of removing the duplicated galaxies in the lightcone on the two-point
clustering measurements. Removing these galaxies will lower the excess clustering
signal on small scales, but it is not guaranteed that this will produce the correct
clustering. This is because for every duplicated galaxy, there is also a galaxy that 
does not exist in the lightcone (see Fig.~\ref{fig:lightcone}), which could potentially also have an effect on the 
measured clustering statistics.

We test the effect of removing the duplicated galaxies on the clustering measurements
using the lightcone built from 9 snapshots.
Pairs of central galaxies in the sample with real-space separation $r<0.3~\hMpc$ are 
identified, and one galaxy in each pair is randomly removed. Due to halo exclusion effects, 
there should be no pairs of central galaxies
with these small separations, and all pairs are due to galaxies being duplicated in the 
lightcone.\footnote{We identified all close pairs in the lightcone, but this could be made faster by only considering galaxies close to the boundary.}

The ratio of the $n(z)$ of the galaxy sample is shown in Fig.~\ref{fig:nz_ratio},
showing the fraction of galaxies that are removed, for the lightcone 
built from 9 snapshots. We denote the redshifts in real and redshift space as
$z_\mathrm{cos}$ (cosmological redshift) and $z_\mathrm{obs}$ (observed redshift), respectively.
When the shells in the lightcone are cut based on the real-space galaxy positions, the
reduction in the real-space $n(z_\mathrm{cos})$ can only be seen in very narrow bins at 
the redshift of each boundary (the black curve). In redshift space, the dips in $n(z_\mathrm{obs})$ are broadened by the effect of velocities. 
The effect is the strongest when the shells are cut based on the redshift-space positions of galaxies (in blue), where the dips in $n(z_\mathrm{cos})$ are much deeper and 
broader than in real space. If the snapshots are instead cut into shells based on the
real-space galaxy positions, with the velocities applied afterwards, the effect is much smaller (in red). In this case,
the dips are also blue-shifted to lower redshifts than the boundaries
in the lightcone, since the velocity of the duplicated galaxies is always towards the
observer (see Fig.~\ref{fig:lightcone}).

After randomly removing one member of each pair, we re-calculate the galaxy clustering.
The random catalogue is generated by randomly sampling redshifts from the galaxies in the
lightcone, and assigning random right ascension and declination coordinates that are uniformly
distributed on the sky. A new random catalogue is generated after removing the duplicated
galaxies, in order to take into account the small change in the $n(z)$.\footnote{We have checked that if the original random catalogue is used, the effect
on the clustering measurements is small.}

The small-scale clustering is shown in Fig.~\ref{fig:xi_small_real},
for the lightcones with different numbers of snapshots. These measurements are the
same as shown in Fig.~\ref{fig:xi_shells_real}, but without rescaling $\xi$ by
any factors of $r$ to better see the differences between the curves
on small scales.
For the 9-snapshot mock, we show the clustering with and without duplicated
galaxies included (in red and purple, respectively).
The monopole is shown in the top left panel, with the difference in the bottom
left, relative to the mock constructed from a single snapshot.
Below $\sim 0.3~\hMpc$, the monopole approaches $\xi_0=-1$ for the mock
built from 1 snapshot, showing that there are almost no 
pairs on these scales, due to halo exclusion effects. As
more snapshots are included, the clustering gets stronger
and stronger on small scales, and is strongest for the mock
built using all 9 snapshots. When the duplicated galaxies are removed, this 
reduces the clustering signal below $0.3~\hMpc$, bringing the monopole
into agreement with the single-snapshot lightcone. The right panel
of Fig.~\ref{fig:xi_small_real} shows the quadrupole.
On small scales, there is a non-zero signal due to the duplicated
pairs, which is the strongest for the mock with 9 snapshots. Removing
the duplicates also removes this clustering signal, bringing it 
consistent with zero.

The clustering in redshift space is shown in the top of Fig.~\ref{fig:xi_small_redshift}. 
This is for a lightcone where the snapshots were cut into shells based on the
position of each galaxy in redshift space, including the effect of velocities. 
For the monopole in the left-hand panel,
we see a similar trend as in real space, where the clustering
on small scales is strongest for the lightcone with 9 snapshots, but by a smaller
amount than in real space. Removing duplicates reduces the clustering,
bringing it into better agreement with the single snapshot. The same is seen
in the quadrupole in the right panel.

The bottom of Fig.~\ref{fig:xi_small_redshift} shows the clustering
in redshift space again, but for the lightcones where the snapshots
were cut into shells in real space, and the effect of velocities was 
applied after making the lightcone. As was also seen in Section~\ref{sec:clustering},
the clustering of the 9-snapshot lightcone is in much better agreement with the
1-snapshot mock, compared to in real space, or compared to when the lightcones were
constructed in redshift space.
The boost in the monopole is only seen on very small
scales ($\sim 0.1~\hMpc$). The quadrupole also shows better
agreement, but with a small excess on scales below $\sim 1~\hMpc$.
As seen in the other mocks, removing the duplicated galaxies reduces the
excess clustering signals, bringing the measurements into better agreement
with the lightcone constructed from a single snapshot.

\section{Conclusions}
\label{sec:conclusions}

In the analysis of large galaxy surveys, it is essential to rely on realistic mock catalogues in order
to validate theoretical models, and understand how the measurements are affected by
systematics. As current and future galaxy surveys get larger, it is increasingly important to make the 
mocks as accurate as possible, creating lightcones that include redshift evolution. Ideally, lightcone
mocks would be constructed from the lightcone output of a N-body simulation, but for many simulations,
only snapshots outputs in the cubic box are available at discrete times. A common method of making
approximate lightcones from the snapshots outputs is to join them in spherical shells. Making lightcones
this way is computationally easy to do, but has the issue that there are discontinuities at the
boundaries where two snapshots are joined. It is possible for a galaxy to appear twice, and be repeated
at either side of one of the interfaces, or to not appear in the lightcone at all.

We test the accuracy of lightcone mocks constructed from snapshots using 4 all-sky lightcones constructed
from the MXXL simulation. The galaxies in these lightcones are assigned $r$-band magnitudes, to
match the evolving luminosity function measured from the SDSS and GAMA surveys. The lightcones we use are
created using 1, 3, 5 and 9 snapshots, where the snapshots are cut in shells in either real space
or redshift space. We measure the two-point clustering
statistics of central galaxies in a volume limited sample with $z<0.2$ and absolute magnitude
$M_r>-20$.

There is a boost in the monopole on small scales, due to galaxies which are duplicated at the boundaries 
between snapshots. In real space, this effect is larger as more snapshots are included, but is also
shifted to smaller scales. In redshift space, this effect is smoothed out by including the velocities.
It is the smallest in the case that the snapshot are cut into shells in real space, with the
effect of velocities applied afterwards. Similar effects are also seen in the quadrupole on small scales,
which is also boosted by the inclusion of duplicated galaxies. 
The clustering is boosted on physical scales $\lesssim 1~\hMpc$, and this scale depends on the distance that galaxies travel between the two snapshots.

We test the effect of randomly removing duplicated galaxies in the 
9-snapshot lightcone on the two-point clustering measurements. 
This is done by identifying all pairs with a real-space separation 
$r < 0.3~\hMpc$, and randomly removing one galaxy in each pair.
On these scales there are no genuine pairs, due to halo
exclusion effects. Both in real space and redshift space, this
is able to reduce the excess small-scale clustering signal.

In this paper, we focus on central galaxies only, where the
effect of duplicated galaxies is the strongest. Including satellites
will reduce this, since the 1-halo term dominates on small scales,
and most pairs come from satellites within the same halo. 
However, there will also be some satellites that are duplicated at the boundaries in the lightcone. 
The impact of including satellites depends a lot on the galaxy sample, e.g. LRGs contain very few satellites.
For the galaxy sample used in this paper, which has a 28\% satellite fraction, we have checked the impact of including satellites in real space. While the effect is smaller than when only centrals are used, there is still some excess small-scale clustering which is at a level greater than $1\sigma$. 
When satellites are included,
the same method can be used as before to identify duplicated central galaxies, but the randomly
removed central galaxy would also have its satellites removed.

To summarise, in order to create lightcone mocks by joining snapshots in spherical shells, we propose 
using all available snapshots, joining them together in shells based on the real-space positions of
galaxies. Galaxies that appear twice in the lightcone can be removed by identifying close
pairs of centrals galaxies at the boundaries. However this does not take into account the galaxies
that are missing in the lightcone. If each shell in the lightcone is made slightly wider, so
that there is a small overlap in the volume at each boundary, the missing galaxies would be 
included in the lightcone. This would have the effect of increasing the number of spurious duplicates, but the same method we have employed can be used to remove them.

This only affects very small scale clustering statistics, below $\sim 1~\hMpc$. Large scales are unaffected, so for many applications of lightcone mocks, such as a BAO analysis, no correction is necessary. However for other applications where the small-scale clustering is important, such as a joint cosmology and HOD analysis, the results could potentially be affected by this systematic. Other applications, such as assessing the impact of fibre collisions, will also be affected by the spurious pairs of repeated galaxies.

\section*{Acknowledgements}

AS would like to thank Francisco Prada, Anatoly Klypin and Shadab Alam for helpful discussions.
This work used the DiRAC@Durham facility managed by the Institute for Computational Cosmology on behalf of the STFC DiRAC HPC Facility (www.dirac.ac.uk). The equipment was funded by BEIS capital funding via STFC capital grants ST/K00042X/1, ST/P002293/1, ST/R002371/1 and ST/S002502/1, Durham University and STFC operations grant ST/R000832/1. DiRAC is part of the National e-Infrastructure.

\section*{Data Availability}

The lightcone mocks underlying this article will be shared on reasonable request to the corresponding author.



\bibliographystyle{mnras}
\bibliography{ref} 








\bsp	
\label{lastpage}
\end{document}